\documentclass[11pt]{amsart}
\usepackage{amsmath,amsfonts,amsthm,amssymb,eufrak,comment}
\usepackage{mathrsfs}
\usepackage{amscd}
\usepackage{fancyhdr}
\usepackage{array}
\usepackage{xy}
\usepackage{epsfig}\usepackage{graphicx}
\usepackage{color}
\usepackage{mathtools}

\input xy

\xyoption{all}

\usepackage{euscript}

\usepackage{footnote}
\makesavenoteenv{tabular}
\makesavenoteenv{table}

\newcounter{thecounter}
\numberwithin{thecounter}{section}

\newtheorem*{lemma*}{Lemma}

\newtheorem{theorem}[thecounter]{Theorem}

\newtheorem{corollary}[thecounter]{Corollary}

\DeclareMathOperator{\SL}{SL}

\DeclareMathOperator{\Aut}{Aut}

\DeclareMathOperator{\ad}{ad}

\def\im{\mathrm{im}}
\def\re{\mathrm{re}}
\renewcommand{\a}{\alpha}

\def\Z{{\mathbb Z}}
\def\R{{\mathbb R}}

\def\N{{\mathbb N}}

\begin{document}

\title{Imaginary reflections and discrete symmetries\\ in the heterotic Monster}

\begin{abstract}

Let $\mathbb{M}$ be  the Monster finite simple group. We give an interpretation of certain discrete symmetries of a family of heterotic string compactifications to $1 + 1$ dimensions in terms of discrete symmetries of the Monster Lie algebra $\frak m$, and more generally Carnahan's family of Monstrous Lie algebras $\frak m_g$, for $g\in\mathbb{M}$ of Fricke type. We relate a Weyl group-type reflection $w_{\im}$, with respect to an imaginary simple root, to a composition of T-duality, time reversal, and parity reversal transformations in the compactified heterotic string. The transformation $w_{\im}$ also has a natural permutation action on the BPS states of the compactified heterotic string.

\end{abstract}


\author{Lisa Carbone}
\address{Department of Mathematics, Rutgers University, Piscataway, NJ 08854-8019, USA}
\email{lisa.carbone@rutgers.edu\footnote{Corresponding author}}

\author{Natalie M. Paquette}
\address{Department of Physics, 
University of Washington,
Seattle, WA 98195-1560}
\email{npaquett@uw.edu}

\thanks{The first author's research is partially supported by the Simons Foundation, Mathematics and Physical Sciences-Collaboration Grants for Mathematicians, Award Number 422182. The second author's research is supported by the University of Washington and the DOE award DE-SC0022347.}

\maketitle

\section{Introduction}

This work concerns the realization of discrete symmetries of a certain compactification of the $E_8\times E_8$- heterotic string as discrete symmetries of the Monster Lie algebra.

Let $\mathbb{M}$ be the Monster finite simple group. Borcherds (\cite{B1}--\cite{B5}) constructed the Monster Lie algebra $\frak m$   to prove part of the Conway--Norton Monstrous Moonshine Conjecture (\cite{CN}).  A fundamental component of Borcherds' construction was Frenkel, Lepowsky and Meurman's (\cite{FLM}) Moonshine Module $V^\natural$,  a graded $\mathbb{M}$-module with $\Aut(V^\natural)=\mathbb{M}$. Some of the tools used in Borcherds' proof were explicitly inspired by conformal field theory and string theory.

The Monster Lie algebra $\frak m$ is  a   quotient of the `physical space'  of the vertex operator algebra $V=V^\natural\otimes V_{1,1}$, where $V^\natural$ is the Moonshine Module of \cite{FLM} and $V_{1,1}$ is  a vertex operator algebra  for the even unimodular  2-dimensional Lorentzian lattice  $II_{1,1}$. 
The Monster Lie algebra $\frak m$ also has a realization as a Borcherds (generalized) Kac--Moody algebra in terms of generators and relations (Section~\ref{MLA}).

Recently, Paquette, Persson, and  Volpato  (\cite{PPV1}, \cite{PPV2})  proposed a  physical interpretation of the  `Monstrous Moonshine Conjecture' 
in terms of a certain family of compactifications, labeled by conjugacy classes of $\mathbb{M}$, of the $E_8\times E_8$ heterotic string to $1 + 1$ dimensions. 

Norton (\cite{No1}) generalized the Monstrous Moonshine Conjecture. The Generalized Moonshine Conjecture  includes a family of irreducible $g$-twisted modules $V^\natural (g)$, one  for each $g \in \mathbb{M}$. Each $V^\natural (g)$ is  a certain graded  representation
 of the centralizer $C_{\mathbb{M}}(g)$. When $g = 1$,  $V^\natural$ is the Moonshine Module of \cite{FLM}.
Dong, Li and Mason (\cite{DLM1}, \cite{DLM2}) and Carnahan (\cite{C1}--\cite{C4}) solved aspects of the Generalized Moonshine Conjecture.

As part of this work, Carnahan (\cite{C1}--\cite{C4}) constructed a family of Lie algebras $\mathfrak m_g$, one 
 for each conjugacy class $[g] \in \mathbb{M}$, associated to the irreducible $g$-twisted modules
$V^\natural (g)$. He showed that the Lie algebras $\mathfrak m_g$ are Borcherds (generalized) Kac--Moody algebras with a real simple root. He used the  twisted denominator identities to prove Norton's Generalized Moonshine Conjecture for
$V^\natural (g)$.

In \cite{PPV1} and \cite{PPV2}, the authors showed that their compactified string models have  Carnahan's Monstrous Lie algebras $\frak m_g$ as  algebras
of spontaneously broken gauge symmetries.

In \cite{CCJMP}, the authors showed that for the Monstrous Lie algebras $\frak m_g$, for $g\in\mathbb{M}$ of Fricke type, there is a  reflection $w_{\im}$ with respect to an imaginary simple root that preserves the bilinear form and root multiplicities.  When composed with reflection in the unique real simple root, this symmetry extends to the Cartan involution on $\frak m_g$. 

This discrete symmetry $w_{\im}$ of the family $\frak m_g$, for $g\in\mathbb{M}$ Fricke, turns out to be related to the compactified string models of \cite{PPV1} and  \cite{PPV2}. In this work (Theorem~\ref{CPT}), we give an interpretation of $w_{\im}$ in terms of a composition of T-duality, time reversal and parity reversal transformations in the compactified heterotic string.


The BPS states of the compactified theory are indexed by pairs $(m,n)$ in the root lattice, thus $w_\im$ has a natural permutation action on them (Section~\ref{BPS}).

The fact that BPS states in string theory form an algebra was first proposed by
Harvey and Moore (\cite{HM1}, \cite{HM2}),
though a complete understanding of the algebra of BPS states of the heterotic string is still missing. 

In future work, we hope to probe this question further by constructing a Chevalley group analog $G(M(\rho))$ for the Monster Lie algebra $\frak m$ in terms of a module $M(\rho)$ whose highest weight is the Weyl vector $\rho$ of $\frak m$. Such a  group  will have a natural action on the module $\bigwedge^\infty \frak n^-$ realized by BPS states in the compactified heterotic string, though a physical interpretation of this group action is currently not known.

The authors would like to thank Scott H. Murray and Roberto Volpato for helpful comments and for careful reading of the paper.

\section{The Monster Lie algebra}\label{MLA}

Let $\mathbb{M}$ denote the Monster finite simple group.  The Monster Lie algebra $\frak m$  has a realization as the Borcherds (generalized) Kac--Moody algebra $\mathfrak m =\mathfrak g(A) / \mathfrak z$ where  $\mathfrak g(A)$ is the Lie algebra  with infinite generalized Cartan matrix $A$ and  $\frak z$ is the center of $\frak g(A)$ (\cite{Jur1}, \cite{Jur2}, \cite{B1}--\cite{B5}):

\begin{figure}[h]\begin{center}
	
		\includegraphics[scale=0.29]{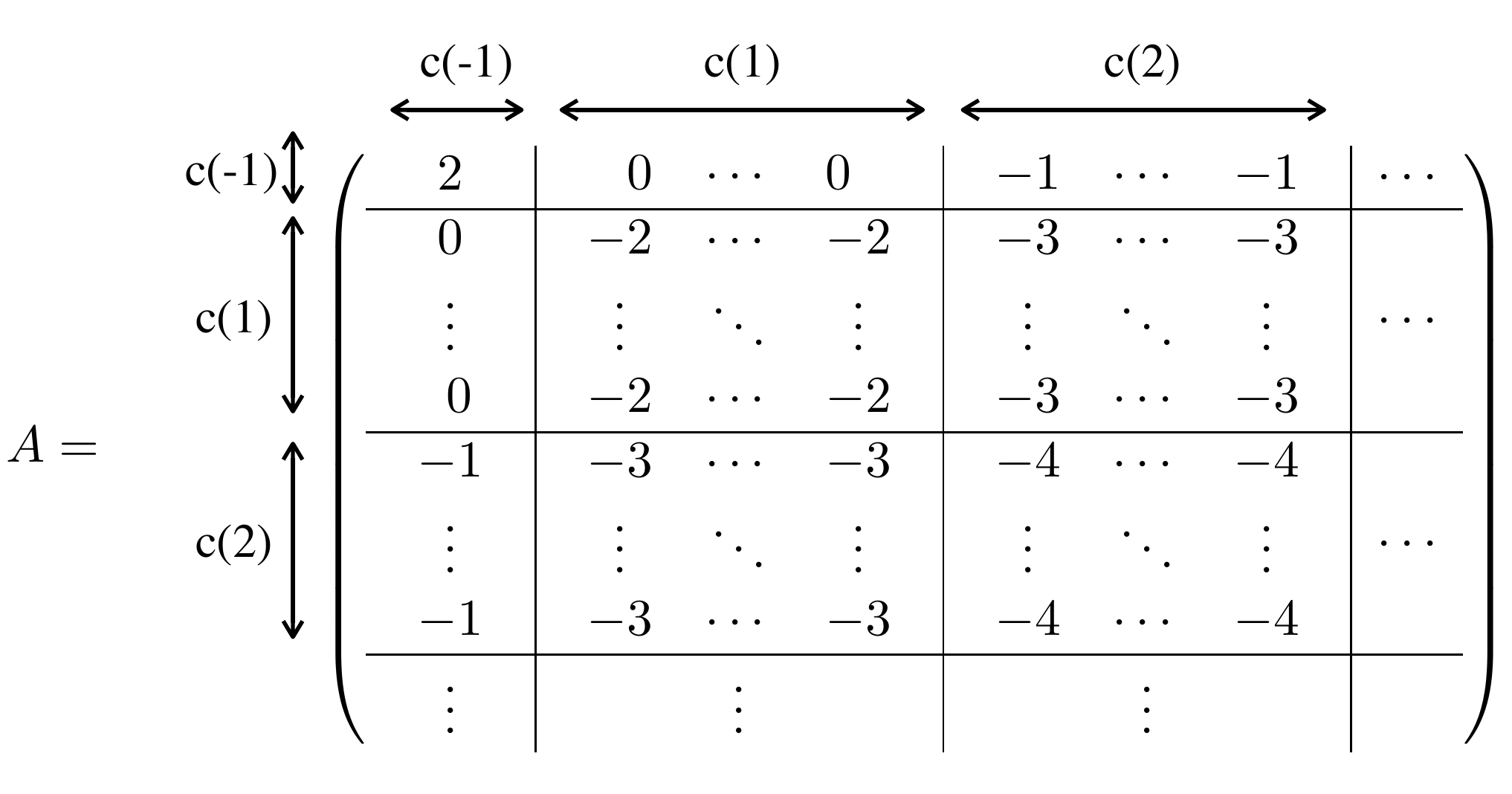}
		\end{center}
	\end{figure}
	
 The numbers $c(j)$ are coefficients of $q^j$ in the  modular function $J(q)=j(q) -744=$
$$\sum_{i\geq -1}c(j)q^j= \frac1q + 196884 q + 21493760 q^2 + 864299970 q^3  + \cdots$$
so $c(-1) = 1$, $c(0) = 0$, $c(1) = 196884$, $\dots$.


\subsection{Generators and defining relations  for $\mathfrak{m}$}

 For $j\in\Z$, we recall the definition of  $c(j)$ above. Define index sets 
\begin{align*}I^{\im} &:= \{(j,k)\mid j\in\N,\; 1 \leq k\leq c(j)\},\\
I &:= \{(j,k)\mid j\in\Z\} \\
&= \{(-1,1)\} \sqcup I^{\im}.
\end{align*}
 The Lie algebra $\mathfrak{g}(A)$  has generating set
 $\{{e}_{jk},\ {f}_{jk},\ h_{jk}\mid (j,k)\in {I}\} $
and defining relations 
\begin{align*}
 [h_{jk},h_{pq}]&=0,\\
 [h_{jk},{e}_{pq}]&=-(j+p) {e}_{pq},\\
 [h_{jk},{f}_{pq}]&=(j+p) {f}_{pq},\\
 [{e}_{jk},{f}_{pq}]&=\delta_{jp}\delta_{kq}h_{jk},\\ 
 (\ad {e}_{-1\,1})^j \,{e}_{jk}&= (\ad {f}_{-1\,1})^j \,{f}_{jk}=0,
\end{align*}
for all $(j,k),\,(p,q) \in {I}$. 

 The $h_{ik}$ are linearly dependent, so $\mathfrak m$ has a two dimensional Cartan subalgebra $\frak h$ with  basis elements denoted $h_{1},$ and  $h_{2}$ (\cite{Jur1}, \cite{Jur2}).  As usual, $\frak h^*$ denotes the dual space of $\frak h$.

The Monstrous Lie algebras $\frak m_g$ are also given by generators and relations associated to Cartan matrices which may be of two possible types (see \cite{C2}, Section 4).

\subsection{Root lattice $II_{1,1}$ for $\frak m$}

We may identify the root lattice of $\mathfrak{m}$ with the Lorentzian lattice $II_{1,1}$ which is $\Z\oplus\Z$ equipped with the bilinear form given by the matrix 
 $\left(\begin{smallmatrix}  ~0 & -1\\-1 & ~0
 \end{smallmatrix}\right)$. Thus roots of $\frak m$ may be denoted by $(m, n)\in II_{1,1}$.

\begin{figure}
\includegraphics[height=5.8cm]{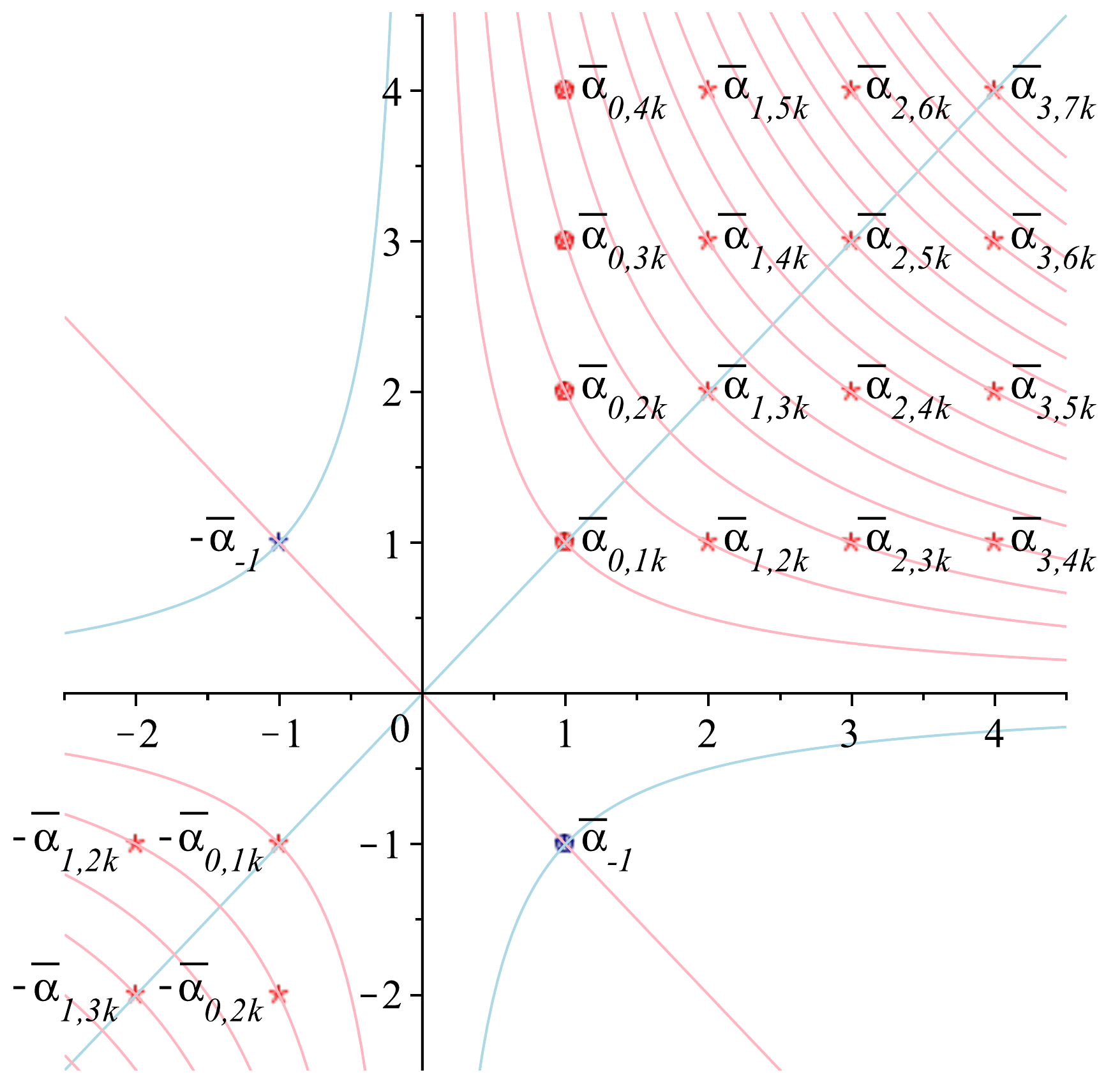} 
\caption{The root system of $\frak m$ in $II_{1,1}$}\label{monsterroots}
 \end{figure}

The bilinear form $(\cdot,\cdot)$ on $\frak m$ defined by the matrix $A$ satisfies
$$ (\a_{jk},\a_{pq})=a_{jk,pq} = -(j+p) $$
for $(j,k),\,(p,q)\in {I}$.
Hence $\a_{-1}=(-1,1)$ is a real simple root with squared norm $2$, while $\alpha_{0,jk}=\a_{jk}=(1,j)$ is an imaginary simple root
with squared norm $-2j$ for $(j,k)\in I^{\im}$. 

 We let $\Delta$ denote the set of roots of $\frak m$. In Figure~\ref{monsterroots}, real roots are denoted by blue nodes on the blue hyperbola with squared norm 2,  imaginary  roots are denoted by red nodes on the red hyperbolas with squared norms $\leq -2$. 

The Lie algebra $\frak m$  has the usual triangular decomposition
$$\frak m = \frak n^-\oplus \frak h \oplus\frak n^+$$
where  
$\frak n^{\pm} =\bigoplus_{\alpha\in \Delta^{\pm}}\frak m_{\alpha}$, $\frak h$ is the Cartan subalgebra, $\Delta^\pm$ are the sets of positive (respectively negative) roots and  $\frak m_\a$ are the root spaces.

For a root $\alpha\in\Delta$, the dimension of the root space $\frak m_\alpha$ is called the {\it multiplicity} of $\alpha$. The imaginary root $(m, n)$ has multiplicity $c(mn)$ (\cite{B1}--\cite{B5}), where the numbers $c(j)$ are coefficients of $q^j$ in the  modular function $j(q)$.

 The Weyl transformation $w_{\re}$ on $II_{1,1}$ with respect to the unique positive real root $\alpha_{-1}$
is $w_{\re}:(n, m) \mapsto (m, n)$.

\subsection{Root system of $\mathfrak m_g$}

 We recall that the McKay-Thompson series labeled by an element $g\in \mathbb{M}$ of order $N$ is
$$T_{1,g}(\tau)=\sum_{n=1}^\infty Tr_{V^\natural_n} (g) q^n\ ,\quad q=e^{2\pi i\tau}\ 
$$

 The algebra $\mathfrak m_g$ is graded by $II_{1,1}(-1/N)\cong \Z\oplus \frac{1}{N}\Z$ where $N\in\Z_{\geq 1}$ with inner product $((a, b/N),(c, d/N)) =
ad+bc/N$ for all $a,b,c,d\in\Z$.

 Recall that an element $g\in\mathbb{M}$ is {\it Fricke} if and only if the McKay-Thompson series $T_{1,g}(\tau)$ is invariant
under the level $N$ Fricke involution $\tau \mapsto -1/N\tau$ for some $N$.

If $g$ is non-Fricke, (which occurs when  $T_{1,g}(\tau)$ is regular at zero) then $\frak m_g$ has norm zero simple roots and no real simple roots. For the purposes of studying certain discrete symmetries, we will omit this case from our work and assume from now on that $g\in\mathbb{M}$ is Fricke.

 If $g$ is Fricke, the positive roots of $\frak m_g$ are $(m,\frac{n}{N})$ for $m,n>0$ which have norm $\frac{-2mn}{N}$. The unique real simple root is $(1,\frac{-1}{N})$. The imaginary simple roots are $(1,\frac{n}{N})$ with multiplicity $\widehat{c}_{1,n}(\frac{n}{N})$ (defined below). The simple roots are thus  $\{(1,\frac{-1}{N}),  (1, \frac{1}{N}), (1, \frac{2}{N}), \dots\}$. The remaining positive roots $(m,\frac{n}{N})$ are imaginary with multiplicity $\widehat{c}_{m,n}(\frac{mn}{N})$ (defined below) (see \cite{C3}).

When $g\neq 1$ and $N\neq 1$, the  multiplicity $\widehat{c}_{m,n}(\frac{mn}{N})$ is the Fourier coefficient of $F_{m,n}$, the (discrete Fourier transform of the) generalized  moonshine functions $\widehat{F}_{i, j}$, where
$${F}_{i, k}=\frac{1}{N}\sum_{j\in\Z/N\Z}e^{-jk/N}\widehat{F}_{i,j}(\tau),\quad \widehat{F}_{i, j}=\sum_{k\in\Z/N\Z}e^{jk/N}F_{i,k}(\tau)$$
(see eq.(4.22), \cite{PPV1} and pg. 7 \cite{C3}). The ${F}_{i, k}$ are the generating functions for the graded dimensions of the spaces $V^\natural_{i,k}$ (see eq.(3.20), \cite{PPV1}).

 When $g=1$ and $N=1$, the multiplicity $c(mn)$ is the coefficient of $q^j$  in the  modular function $J(q)$ for $j=mn$.

In this paper, we will focus on elements $g$ of order $N$ with a trivial multiplier, for simplicity. Recall that for any McKay-Thompson series 
$T_{1,g}(\tau)$
labeled by an element $g\in \mathbb{M}$ of order $N$, a multiplier is a nontrivial phase $\xi_g: \Gamma_0(N) \rightarrow U(1)$, in particular an $N$-th root of unity, that the function may acquire under a modular transformation by an element of $\Gamma_0(N) \leq\SL(2,\mathbb{Z})$. Here, as usual $\Gamma_0(N) = \left\{ \begin{pmatrix} a & b \\ c & d \end{pmatrix} \in\SL(2,{\Z})\mid c \equiv 0 \pmod N \right\}$. Physicists also refer to this multiplier as an `anomaly'. The order of the multiplier\footnote{In the case of the Monster vertex operator algebra, $\lambda_g$ is always a divisor of 24 for any $g$. } is denoted by $\lambda_g$, so that $\xi_g^{\lambda_g}=1$, where $\lambda_g \vert N$. In the language of vertex operator algebras, having a trivial multiplier means that the corresponding orbifold $V^{\natural}/ \langle g \rangle$ is non-anomalous. In general, the orbifolds may have a nontrivial multiplier of order $\lambda_g$, whereby $V^{\natural}/\langle g^{\lambda_g}\rangle$ is non-anomalous. We refer to \cite{PPV1} for details. In this case, the root lattices of $\mathfrak m_g$ are spanned by vectors of the form
\begin{equation}
\begin{pmatrix} m \\ n \end{pmatrix} = k \begin{pmatrix} \lambda_g \\ 0 \end{pmatrix} + \ell \begin{pmatrix} \mathfrak{E}_g \\ {1 \over N \lambda_g} \end{pmatrix},
\end{equation} 
where $k, \ell \in \mathbb{Z}, \mathfrak{E}_g \in \mathbb{Z}/\lambda_g\mathbb{Z}$, and $\mathfrak{E}_g$ is coprime to $\lambda_g$. 

\subsection{Imaginary reflection}

 In  [CCJMP], the authors, with Chen, Jurisich and Murray, showed that  there is a candidate for a Weyl reflection of $\frak m_g$, for $g\in\mathbb{M}$ Fricke, in terms of an imaginary simple root.

\begin{theorem}\label{imagref} ([CCJMP]) Let $g\in\mathbb{M}$ be Fricke. Let $\frak h$ be the Cartan subalgebra of  $\frak m_g$. For $\alpha_{\im}=(1, \frac{1}{N})\in II_{1,1}(-1/N)$, there is a map $w_{\im}:\frak h^\ast\to\frak h^\ast$  with the properties
	\begin{enumerate}
		\item $w_{\im}(m,n) = (-n,-m)$
		\item $w_{\im}$ has order 2.
		\item $w_{\im}$ preserves the bilinear form $(\cdot,\cdot)$. That is  $(\alpha,\beta)=(w_{\im}(\alpha),w_{\im}(\beta))$ for all $\alpha,\beta\in\Delta$.
		\item $w_{\im}$ preserves root multiplicities. That is, for all $m,n\in\Z$
$$mult(w_{\im} \cdot \frak m_{g,(m,n)})=mult(\frak m_{g,(-n,-m)})=\widehat{c}_{-n,-m}(\frac{nm}{N})=\widehat{c}_{m,n}(\frac{mn}{N}).$$

	\end{enumerate}
\end{theorem}

The proof of (4) in Theorem~\ref{imagref} is easily seen when $g=1$ and $N=1$, as in this case, $c(mn)=c(nm)$, but is non trivial for $g\neq 1$ (see \cite{CCJMP}).

The following theorem shows that the composition of the real Weyl reflection $w_{\re}$ and imaginary reflection $w_{\im}$ coincides with the Cartan involution on $\frak m_g$.

\begin{theorem} ([CCJMP]) Let $g\in\mathbb{M}$ be Fricke. 
	 The map $\omega:=w_{\im} w_{\re}=w_{\re}w_{\im}$ extends to an involution on $\frak m_g$ satisfying
	$\omega({\frak m}_{g,\alpha})\subseteq{\frak m}_{g,-\alpha}$ for all $\a\in\Delta$ and $\omega(h) =-h$ for all $h\in \frak h$.
	\end{theorem}


\section{Monstrous Moonshine and the heterotic string}

In \cite{PPV1}, the second author, with Persson and Volpato, proposed a physical interpretation of the Monstrous Moonshine Conjecture in the context of a family of compactifications of the $E_8\times E_8$ heterotic string to $1+1$ spacetime dimensions. We refer the reader to \cite{PPV1} for details. The family of compactifications is labeled by conjugacy classes of the Monster, which we denote by a generating element $g$. See also \cite{PPV2} for an introduction and summary to the work \cite{PPV1}, as well as some key results phrased in the language of vertex operator algebras.

Each such compactification realizes Carnahan's Monstrous Lie algebra $\frak m_g$ as an algebra of spontaneously broken gauge symmetries which generate the spectrum of BPS states in $1+1$ dimensions (\cite{PPV1}). The compactifications are constructed using the \cite{FLM} Moonshine Module $V^\natural$ (for $g=1$) and orbifolds  $V^{\natural}/ \langle g \rangle$ thereof, for $g \neq 1$.\footnote{See \cite{PPV1} for details on how to define a consistent string compactification even when the orbifold vertex operator algebra $V^{\natural}/\langle g \rangle$ is anomalous.}.

The $1+1$-dimensional spacetime has metric of Lorentzian signature and the topology of a cylinder $S^1\times\R$, where $\R$ is the timelike direction and $S^1$ is a spacelike circle of radius $R$ when $g=1$. This compactification radius $R$ plays a role in the results that follow. When $g \neq 1$ and the orbifold $V^{\natural}/\langle g \rangle$ is non-anomalous, one considers an orbifold of the $S^1$ by the generator of translations by $1/N$-th of the period of the $S^1$; when the orbifold vertex operator algebra  is anomalous, one considers an orbifold by a ${1 \over N \lambda_g}$ fraction of the period with $\lambda_g > 1$ (see \cite{PPV1} for details). 

The number of times a closed string winds around the cylinder is called the {\it winding number} $w$. We use $m$ to denote the {\it momentum} of the closed string. The winding number $w$ and momenta $m$ satisfy quantization conditions and hence pairs $(w,m)$ live on a lattice \footnote{For the $\lambda_g >1$ case, one defines $w:= {\ell \over N \lambda}, m:= k \lambda - \ell \mathfrak{E}_g$, and $N \mapsto N \lambda_g$. We expect our subsequent results generalize, after employing results of \cite{PPV1}, to this case.}.

One can also consider a further compactification of the timelike direction to a circle of radius $\beta$ after performing a Wick rotation to Euclidean signature, so that the spacetime becomes a Euclidean torus $T^2$ (\cite{PPV1}). For each model, the BPS indices below are defined on $T^2$ and we restrict to this Euclidean spacetime geometry in what follows.

\subsection{Discrete symmetries in the compactified string}

Discrete symmetries of the heterotic string models of \cite{PPV1} and \cite{PPV2} on $T^2$ are discrete subgroups of 
$$O(L) \leq O(2,2:\Z) < O(2,2:\R)$$
where $L$ is a 4-dimensional lattice, denoted $\Gamma^{2,2}$, of signature $(2,2)$ consisting of momentum and winding vectors  $(w_1,m_1,w_2,m_2)$ on $T^2$. The groups are presented in Theorems 1 and 2 of [PPV1] for the discrete groups associated to each conjugacy class. 

In the case $g=1$, for each vector in $L$ and fixed compactification radius $R$, is associated a left-moving momentum $k_L \in \R^2 $. One can construct the vertex operators $V_\chi e^{ik_LX_L}$  in $V^\natural\otimes V_{1,1}$ (defined over $\R$) whose zero modes generate $\frak m$ ({\cite{PPV1}). Similar vertex operators may be constructed when $g \neq 1$, and their zero modes generate $\frak m_g$.

Each lattice $L$  has a 2 dimensional sublattice  $\Gamma^{1,1}$ consisting of momentum and winding vectors  $(w_1,m_1)$ which label the BPS states of the theory. The lattice $\Gamma^{1,1}$ can be identified with the root lattice $II_{1,1}$ for the Monster Lie algebra when $g=1$, and the root lattice for $\frak m_g$ in general.

 The group $\Aut(\Gamma^{1,1})$ is in the connected component of the identity in $O(2,2;\R)$. Hence, discrete symmetries of the model that do not belong to this component act on  $\Gamma^{2,2}$ but not on the sublattice $\Gamma^{1,1}$.

The Weyl transformation on $\Gamma^{1,1}$ with respect to the positive real root exchanges the winding and momenta along the spacelike circle: $(w_1,m_1)\to  (m_1,w_1)$ and keeps $w_2$ and $m_2$ fixed.  There is a natural extension to $\Gamma^{2,2}$ which sends $(w_1,m_1,w_2,m_2)\to  (m_1,w_1,m_2,w_2).$

By Theorem~\ref{imagref}, the imaginary reflection  $w_{\im}$ of \cite{CCJMP} on $\Gamma^{1,1}$ exchanges the winding and momenta up to a sign: $w_{\im}(w_1,m_1) = (-m_1,-w_1).$

%
%

The Cartan involution $\omega$ on $\Gamma^{1,1}$  acts as multiplication by $-1$ on the Cartan subalgebra 
$\mathfrak{h}$ of $\mathfrak{m}_g$ and flips the sign of the left-moving momentum, $k_L \mapsto - k_L$ in the vertex operators $V_\chi e^{ik_LX_L}$. Therefore, it acts on the momentum-winding lattice vectors as $\omega(w_1, m_1) = (-w_1, -m_1)$, while swapping the positive and negative roots of $\frak m_g$. This inversion of the left-moving momentum $k_L$ coincides with the involution $\theta$ of \cite{FLM}.

\section{Reflections and discrete symmetries of the heterotic string}

We let $\mathcal{C}$ denote the charge conjugation transformation, $\mathcal{P}$ denote the parity transformation  and $\mathcal{T}$ denote the time reversal transformation in the compactified heterotic string in Euclidean signature.
The parity transformation $\mathcal{P}$ on $\Gamma ^{2,2}$ changes the sign of momentum and winding in the spatial direction:
$$m_1 \to  -m_1,\quad  w_1 \to  -w_1$$ 
and the Euclidean time reversal is a sign flip of the time coordinate (on the spacetime torus).

The following theorem gives a relationship between the real and imaginary Weyl reflections of $\frak m_g$ and 
 $\mathcal{P}$ and $\mathcal{T}$.

\begin{theorem}~\label{CPT} Let $g\in\mathbb{M}$ be Fricke. In  $\frak m_g$, we have
\begin{align*}
\text{Cartan involution } \omega
&\buildrel\text{[CCJMP]}\over{\quad=\quad}w_{\re}w_{\im}\\
&\buildrel\text{[CCJMP]}\over{\quad=\quad} w_{\im}w_{\re}\\
&\buildrel\text{[PPV1]}\over{\quad\leftrightarrow\quad}V_\chi e^{ik_LX_L} \to V_\chi e^{-ik_LX_L}\\
&\quad\leftrightarrow\quad\mathcal{T}  \mathcal{P}\\
&\quad=\quad\mathcal{P}\mathcal{T}
\end{align*}
where $w_{\re}$ and $w_{\im}$ denote the real and imaginary Weyl reflections of $\frak m_g$ respectively.
We recall that in Euclidean signature, the action of $\mathcal{P}\mathcal{T}$ on the underlying spacetime is simply  a rotation by $\pi$.
\end{theorem}
Theorem~\ref{CPT} is a simple application of the CPT theorem in physics (see for example \cite{K}, \cite{GT}). The Cartan involution corresponds to the action of $\mathcal{C}$ on the charges of BPS states in the physical model. Since BPS states are labelled by $(m,n)\in II_{1,1}$, the action of $\mathcal{C}$ corresponds to $(m,n)\mapsto (-m,-n)$.

An easy proof of Theorem~\ref{CPT}  is given in Subsection~\ref{complexvar} in terms of the complex moduli $T$ and $U$. In particular, if we  consider the theory on a Euclidean cylinder or torus, the final line indicates that the action of $\mathcal{P}\mathcal{T}$, when composed with an additional action of $\mathcal{C}$, produces a symmetry on the heterotic models.

Here  $V_\chi$ is a holomorphic vertex operator  of conformal weight $h = 1 + mw$ where $m$, $w$ are the momentum and winding quantum numbers on the circle (\cite{PPV1}). 

\subsection{Physical symmetries in terms of $T$, $U$}\label{complexvar}

Discrete symmetries of the model can also be described in terms of the complex variables $T$ and $U$ (equation 3.14 of \cite{PPV1}), which can be interpreted as the K{\"a}hler and complex structure moduli, respectively, of the spacetime $T^2$ when working in Euclidean signature.

The Weyl reflection acts as 
$w_{\re}:(T ,U)\mapsto  (U , T)$.

The imaginary reflection acts as $w_{\im}: (T, U) \mapsto (-{U}, -{T})$.

The time reversal transformation acts as $\mathcal{T}:(T, U) \mapsto (\bar{T}, \bar{U})$.

The parity transformation acts as $\mathcal{P}: (T, U) \mapsto (-\bar{T}, -\bar{U})$.

From this, we easily see that
$$w_{\re}w_{im}=w_{\im}w_{\re}\leftrightarrow\mathcal{T}\mathcal{P}=\mathcal{P}\mathcal{T}.$$

\subsection{Physical interpretation of $w_{\im}$}

\begin{corollary}The imaginary reflection $w_{\im}$ is  the composition $$w_{\im}\leftrightarrow w_{\re}^{-1}\mathcal{T}  \mathcal{P}\leftrightarrow w_{\re}\mathcal{T}  \mathcal{P}\leftrightarrow w_{\re}\mathcal{P}  \mathcal{T}$$ of the real Weyl reflection $w_{\re}$ with time reversal $\mathcal{T}$ and parity transformations $\mathcal{P}$. Further, as explained in \cite{PPV1} and \cite{PPV2}, the real Weyl reflection $w_{\re}$ is interpreted as a T-duality along the spacelike circle of the heterotic model. 
\end{corollary}

In other words, up to parity, the imaginary reflection $w_{\im}$ is a time reversal transformation composed with  T-duality. Note that the real Weyl reflection $w_{\re}$ (i.e. spacelike T-duality) is not in the connected component of the identity in $O(2,2;\R)$, but its composition with time reversal is. In physical language, this was referred to in \cite{PPV1}, \cite{PPV2} as a \textit{self-duality} of the model.  The imaginary reflection $w_{\im}$, which is then obtained by a further parity transformation, therefore lies outside the connected component of the identity in $O(2,2;\R)$.


\section{BPS states}\label{BPS}

 The imaginary reflection $w_{\im}$ of \cite{CCJMP} turns out to have a natural action on  the BPS states of the compactified heterotic string of \cite{PPV1}.
 
 Enumeration, or counting, of BPS states has led to  interactions with the theory of automorphic forms. One way to count the BPS states is to use an {\it index}. This is a function which vanishes on the non-BPS representations but is nonzero and counts the BPS representations. 

 In \cite{PPV1}, the authors  introduced a supersymmetric index $Z$ which counts (with signs) the number
of BPS states of the compactified heterotic string.

The fact that BPS states in string theory form an algebra was first proposed by
Harvey and Moore (\cite{HM1}, \cite{HM2}). In the heterotic string constructions discussed presently, however, \cite{PPV1} and \cite{PPV2} realized the Hilbert space of BPS states $\mathcal H$ as a module for
the Monster Lie algebra $\frak m$ when $g=1$ (the analogous statement for $\frak m_g$ holds when $g \neq 1$), while Harvey and Moore showed that there is an algebraic
structure on $\mathcal H$ itself.

The space of BPS states contributing to the supersymmetric index  in \cite{PPV1} and \cite{PPV2} is naturally
isomorphic to $\bigwedge^\infty\frak n_g^-$.

\subsection{Action on BPS states}

The real and imaginary reflections act in an obvious way on the BPS states, since such states are labeled by pairs $(m,n)$ in the root lattice $II_{1,1}$ (\cite{PPV1}). 



 The real reflection permutes the pair of BPS states labeled by  the real roots $\alpha_1=(1,-1)$ and $-\alpha_1=(-1,1)$:
 $$w_{\re}(1,-1)=(-1,1).$$
The imaginary reflection  permutes a pair of BPS states labeled by imaginary roots up to a sign:
$$w_{\im}(m,n)=(-n,-m).$$

\section{Discrete symmetries of the supersymmetric index $Z(T, U)$}

The supersymmetric index $Z(T, U)$ of  \cite{PPV1} corresponding to $g=1\in\mathbb M$, depending on
 $(T,U)\in\mathbb{H}\times\mathbb{H}$, is given as follows:
$$Z(T,U)=\Bigl(e^{2\pi i (w_0T+m_0U)}\prod_{\substack{m> 0\\w\in \Z}}(1-e^{2\pi i Um}e^{2\pi iTw})^{c(mw)}\Bigr)^{24}.$$
The $c(mw)$ are the coefficients of the modular invariant $j$-function and $w_0, m_0$ are the winding and momentum  of the ground state.  For suitable values of $w_0$, $m_0$, $Z(T,U)$ is (the 24th power of) the denominator identity for the Monster Lie algebra $\frak m$ (\cite{PPV1}). They also defined twisted indices $Z_{g,1}(T,U)$ for $g\neq 1$.


The infinite
product formula obeys the famous identity, proven independently by
Koike, Norton, and Zagier (\cite{B6}):
$$j(T) - j(U) = {1 \over p}  \prod_{\substack{m> 0\\n\in \Z}}(1-p^m q^n)^{c_{mn}}$$
where $p=e^{2 \pi i T}$ and $q=e^{2 \pi i U} $. This gives the `KNZ index':
$$Z(T, U)^{1/24} = j(T) - j(U)$$

 which is an automorphic form under $O(L) \cong O(2, 2;\Z) \cong (\SL(2, \Z) 
\times \SL(2, \Z)) \rtimes \Z/2\Z$. Here, $\Z/2\Z$ is the Weyl group. With respect to the additive side of the denominator identity, the $\SL(2,\Z)$ factors act on $j(T)$ and $j(U)$, respectively, while $\Z/2\Z$ switches them.

The real Weyl reflection $w_{\re}:(T , U)  \mapsto (U,T)$ has matrix representation (as an automorphism of $II_{1,1}$) $\left(\begin{smallmatrix} 0 &1 \\1 & 0 \end{smallmatrix}\right)$ and generates the $\Z/2\Z$ subgroup of $O(L)$. Thus $W=\langle w_{\re}\rangle$
 is a symmetry of $Z(T, U)^{1/24}$ up to a minus sign (and hence a symmetry of $Z(T, U)$:
 
\begin{align*}
Z(T, U)^{1/24} &= (e^{-2 \pi i T} - e^{-2 \pi i U})\prod_{m, n >0}(1 - e^{2 \pi i U m}e^{2 \pi i T n})^{c(mn)} \\
&= -(e^{-2 \pi i U} - e^{-2 \pi i T})\prod_{m, n >0}(1 - e^{2 \pi i U m}e^{2 \pi i T n})^{c(mn)} \\
&=-(e^{-2 \pi i U} - e^{-2 \pi i T})\prod_{m, n >0}(1 - e^{2 \pi i T m}e^{2 \pi i U n})^{c(mn)}\\
&= -Z(U, T)^{1/24}.
\end{align*}

The imaginary reflection $w_{\im}:(T , U)  \mapsto (-U,-T)$ has matrix representation $\left(\begin{smallmatrix} ~0 &-1 \\-1 & ~0 \end{smallmatrix}\right)$ (as an automorphism of $II_{1,1}$). The transformation $w_\im$ is not a symmetry of the KNZ index since it is not an element of $O(2,2;\Z)$.





\end{document}